\documentclass[11pt]{article}
\usepackage{graphicx,color} 
\usepackage{amsmath,mathrsfs}

\usepackage{wrapfig}

\usepackage[english]{babel}
\usepackage[T1]{fontenc}

\renewcommand{\theenumi}{(\kern -0.15ex{\roman{enumi}})}

\begin{document}
%\selectlanguage{english}
\onecolumn

\noindent \textbf{\LARGE Variations on chaos in physics: from unpredictability to universal laws} \\[1cm]
Amaury Mouchet \\[2cm]
\let\thefootnote\relax\footnote{Amaury Mouchet\\
Laboratoire de Math\'ematiques et de Physique Th\'eorique, Universit\'e  Fran\c{c}ois Rabelais de Tours\\ 37200 Tours, France}

\newcommand{\thefootnote}{\arabic{footnote}}
\setcounter{footnote}{0}

%\begin{quotation}Any classification is superior to chaos and even a 
%classification at the level of sensible properties is a step 
%towards rational ordering. It is legitimate, in classifying
%fruits into relatively heavy and relatively light, to begin by
%separating the apples from the pears even though shape,
%colour and taste are unconnected with weight and volume.
%This is because the larger apples are easier to distinguish
%from the smaller if the apples are not still mixed with fruit
%of different features. This example already shows that
%classification has its advantages even at the level of aesthetic perception
%\cite[chap.~1]{LeviStrauss68a}.
%\end{quotation}

The tremendous popular success of \textit{Chaos Theory} shares some
common points with the not less fortunate \textit{Relativity}: they
both rely on a misunderstanding. Indeed, ironically, the scientific
meaning of these terms for mathematicians and physicists is quite
opposite to the one most people have in mind and are attracted by. One
may suspect that part of the psychological roots of this seductive
appeal relies in the fact that with these ambiguous names, together with
some superficial clich\'es or slogans immediately related to them
(``the butterfly effect'' or ``everything is relative''), some have
the more or less secret hope to find matter that would undermine two
pillars of science, namely its ability to predict and to bring out a
universal objectivity.

As noted by Planck\footnote{As soon as 1910, Felix Klein had already
 wrote that ``theory of relativity is the theory of invariants of the
 four dimensional space-time domain with respect to a particular
  group of collineations, more precisely, to the Lorentz group''
  \cite[\S~3.1 p.~70]{KosmannSchwarzbach10a}. }
\begin{quotation}
 As a matter of fact the concept of relativity is based on a more
 fundamental absolute than the erroneously assumed absolute which it
 has supplanted \cite[chap.~VI, p.~195]{Planck32a}.
\end{quotation}
A position advocated of course by Einstein himself all along his
writings, for instance
\begin{quotation}
The belief in an external world independent of the perceiving subject is the basis of all natural science \cite[p.~66]{Einstein31a}.
\end{quotation}
I will not say much about Relativity and more generally on how the
notion of symmetry reveals the objectivity of the world; this has been
the subject of my contribution to the previous volume of this series
\cite{Mouchet15a}\nocite{Emmer+15a}. Here I propose to focus on Chaos
Theory and illustrate on several examples how, very much like
Relativity, it strengthens the position it seems to contend with at
first sight: the failure of predictability can be overcome and leads
to precise, stable and even more universal predictions.

Far before it became a scientific notion, chaos describes an absence
of structure, an unorganised confusion. It is probably for
encapsulating this amorphous state of matter that the word ``gas'' was
coined, with still an alchemical flavour, from the latin word
``chaos'' by the flamish Jan Baptist van Helmont in the middle of the
\textsc{xvii}$^{\text{th}}$ Century \cite[pp.~67-69]{Pagel82a}.  The
greek word $\chi\acute\alpha o\varsigma$ itself comes from an old
Indo-European root \textit{ghen} or \textit{ghei} meaning a lack, a
gap or a vacuum. Bearing in mind this meaning, still overspread in
everyday life, talking about a theory, or about laws of chaos seems a
self-contradiction from the very beginning.

In the first section I will start with a simple example that allows to
understand what is meant by chaotic in the context of dynamical
systems. In \S~\ref{sec:chaoseverywhere}, I will explain why chaotic
behaviours are met everywhere for any realistic physical systems.
Then, before I conclude with a historical comment on the butterfly
effect, in \S~\ref{sec:statistics}, I will illustrate how, despite the
unpredictability of one individual evolution, we can nevertheless
establish stable probabilistic properties and laws that emerge from a
collective behaviour.

\section{A simple example of physical chaotic system}

\begin{figure}[!hb]
\begin{center}
\includegraphics[width=10cm]{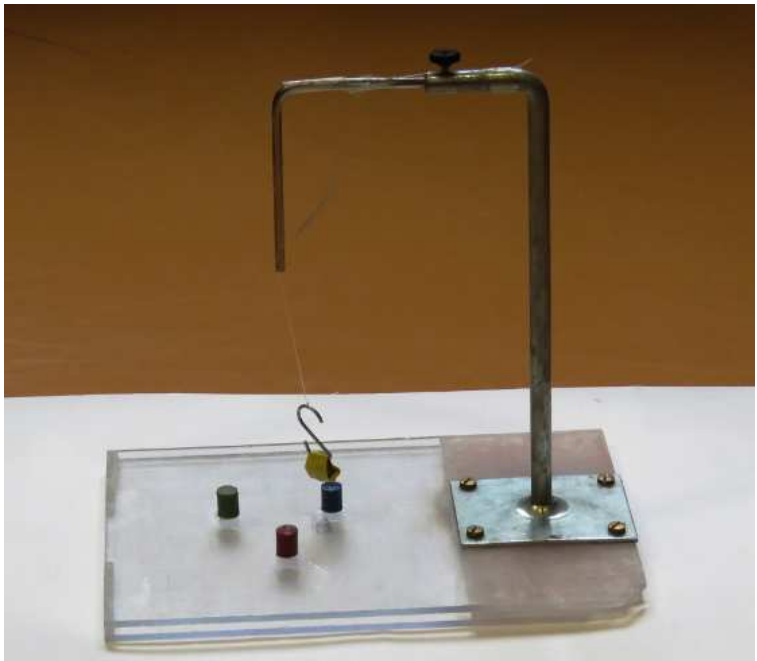}
\caption{\label{fig:pendulemag} The magnetic pendulum is made of a
  magnet attracted by three other identical magnets, distinguished by a
  colour, which are fixed at the vertices of an equilateral triangle
  (here, one edge is 6~cm long).  The constant length of the thread is chosen
  in order to avoid any contact between the magnets during the motion.
  This photography shows the stable equilibrium point~$M_3$ near the
  blue magnet.  }
\end{center}
\end{figure}
\begin{figure}[!ht]\begin{center}
\includegraphics[width=7.5cm,height=6cm,angle=-29]{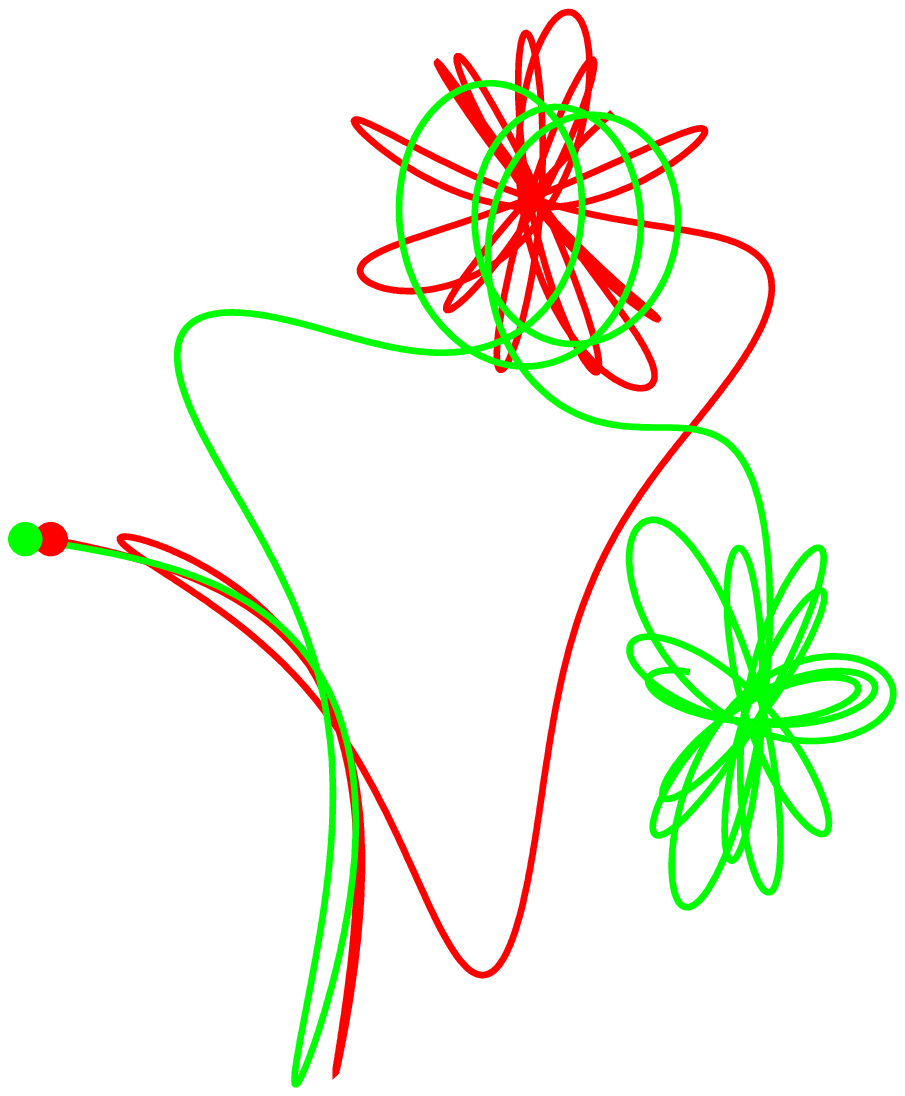}
\caption{\label{fig:pendmag_traj12} Seen from above are shown the
  numerical simulation of two trajectories of the magnetic pendulum:
  the attractive force between the pendulum and the fixed magnet
  separated by distance~$r$ is taken to be proportionnal
  to~$-\vec{r}/r^3$ (Coulombian-like); in addition there is the usual
  linear force due to gravity and directed towards the center of the
  triangle (the stable equilibrium point when there is no magnet) and
  the friction force is taken to be opposite to the speed).  Starting
  with no initial speed from two close initial positions on the left,
  the two trajectories diverge one from each other very quickly (after
  a small fraction of second, which is the natural period of the
  pendulum without magnets) and end at two different equilibrium
  positions. The one stopping at~$M_1$ (resp.~$M_2$) is represented in
  red (resp. green) and a red (resp. green) spot indicates its
  starting point.  }
\end{center}
\end{figure}

A usual simple frictionless pendulum (a mass attached to an
unstretchable thread) has a regular motion : the trajectories of small
amplitudes follow an ellipse.  If we now take the mass to be a magnet
and if, in addition to the gravity field, we create an external
magnetic field with some other fixed magnets, we generically get an
irregular motion, even if the field itself seems to be well-ordered
with some symmetrical structures. For instance, three attractive
identical magnets can be placed at the vertices of an equilateral
triangle whose center lies below the vertical position of the
pendulum. However, if the magnets are strong enough, their attraction
destabilises the vertical position; in fact, there are now three
stable equilibrium positions~$M_1$, $M_2$, $M_3$ inclined towards each
magnet (figure~\ref{fig:pendulemag}). If we let the pendulum evolve
away from these equilibrium positions we observe that the pendulum
follows a rather complicated trajectory passing through the
neighbourhood of~$M_1$, $M_2$, $M_3$ several times before the friction
eventually stops it at one of these three points (figure~\ref{fig:pendmag_traj12}).
\begin{figure}[!ht]
\begin{center}
\includegraphics[width=\textwidth]{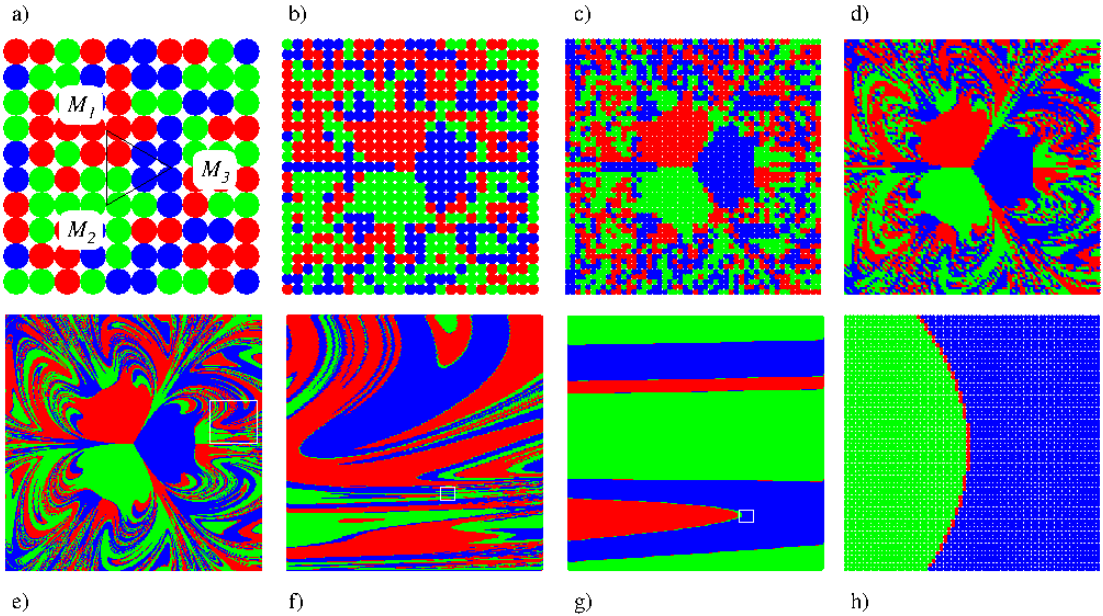}
\caption{\label{fig:pendmag_attracteur} In a) a grid of~$10\times10$
  possible initial positions is chosen and each of them is coloured
  according to the final position the pendulum eventually stops at
  (red for~$M_1$, green for~$M_2$, blue for~$M_3$) with the same model
  as the one used to compute~figure~\ref{fig:pendmag_traj12}. In b),
  c), d) computations are done for a finer grid ($25\times25$,
  $50\times50$ and $100\times100$).  In e)~a~$500\times500$ grid is
  used and the white squares indicate the portion of the picture that
  is zoomed on the immediate right.  At very smale scales like the one
  in h) one can see that the borders always involve the three colours.
}
\end{center}
\end{figure}
Unless we remain in the immediate vicinity of one of the~$M$'s, it
seems impossible to choose the initial conditions (position and speed)
in order to make the pendulum stop at a position chosen in advance. To
illustrate more quantitatively this difficulty, let us first
distinguish the three magnets by a colour. Then associate one of these
colours to each initial position according to the final position of
the pendulum: starting above the point of Cartesian coordinates~$(x,y)$
without initial speed, if the pendulum stops at~$M_1$ (respectively
$M_2$ or~$M_3$) then~$(x,y)$ will be coloured in red (respectively
green or blue).  Experimentally we can only scan a coarse-grained grid
of possible initial positions (say $10\times10$ like in
figure~\ref{fig:pendmag_attracteur}a) but using a computer simulation
of a more or less realistic model one can get a very high resolution
pattern (figure~\ref{fig:pendmag_attracteur}e).  Beyond the uniformely
coloured areas surrounding~$M_1$, $M_2$ and $M_3$, which reflect the
stability of these equilibrium positions, we observe an intricate
fractal-like pattern where all the three colours are intertwined at
arbitrarily small scales (figures~\ref{fig:pendmag_attracteur}e-h). An
infinitesimal shift from a red initial position may fall in a green
area and this means that the two corresponding trajectories of the
pendulum are eventually separated as far as possible one from the
other (final position at $M_2$ instead of $M_1$).  This is an
illustration of the extreme sensitivity of the dynamics with respect
to the initial conditions, or to any kind of perturbation, and this
behaviour characterises what is called a chaotic system.  More
precisely, the linearisation~$\dot{\delta\vec{r}}=A\delta\vec{r}$ of a differential equation in
the neighbourhood of a given solution ---
where~$A$ is a matrix independent of the variation~$\delta \vec{r}$
but generally depends on time~$t$ ---  leads generically to some
exponential behaviour of the
shift~$\delta\vec{r}(t)\sim\delta\vec{r}(0)\,\mathrm{e}^{t/\tau}$
where~$\tau$ is a typical time scale (known as the Lyapunov time).
 
Therefore, it is true that 
predictibility \textit{about one individual trajectory}
 fails beyond durations of order~$\tau$ : it should require
an irrealistic precision on all the experimental conditions or on
all the parameters of the modelisation to predict correctly 
the final position of the pendulum for an arbitrary set of initial conditions. This may  include the gravitational perturbation 
 due to the mass of a thundercloud moving above the place where the pendulum is and, of course,
the famous flap of a butterfly wing in Brazil \cite{Lorenz72a}.

\section{Ubiquity of chaotic behaviour}\label{sec:chaoseverywhere}
 \begin{figure}[!ht]
\begin{center}
\includegraphics[width=\textwidth]{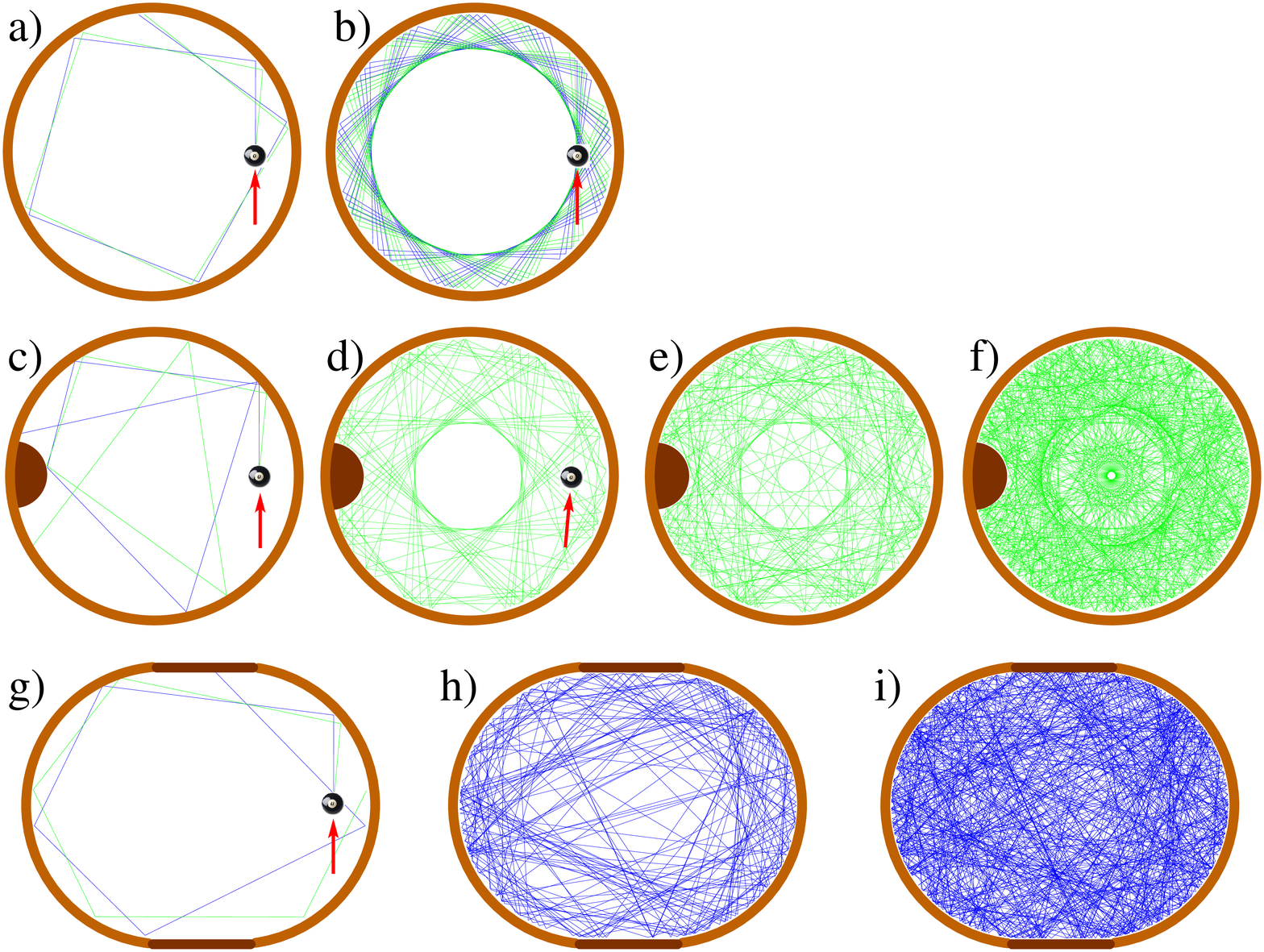}
\caption{\label{fig:billards} A ball bouncing without friction in a circular 2D-billiard has a regular motion.
The difference between two different trajectories, in blue and green in a) and b), is proportional
to the elapsed time, unlike in chaotic systems where this growth is exponential. The two constants of motion (energy and angular momentum)
impose some selection rules that prevent most trajectories to explore the central zone of the billiard.
When the rotational symmetry is broken (by a bump for the billiard in the second line or by stretching the circle into a stadium-shaped billiard in the third line), the number of  constants (only the energy) becomes lower than the
number of degrees of freedom (D=2) 
and the system becomes chaotic: as seen in c) and g) two trajectories split very quickly and, as time grows,
an individual trajectory explores all the area of the billiard. 
From~\cite[figure 10.7]{Mouchet13a}.
}
\end{center}
\end{figure}
As we have seen with the magnetic pendulum there is no need to
consider a system as complex as the Earth atmosphere to deal with a chaotic
system (by the way, Lorenz'climatic model was described by three
parameters \cite[eqs.~(25), (26), (27)]{Lorenz63a}).  In fact chaos is
naturally the rule and regular motion is the exception: when
dissipation is negligible, for a system with~$D$ degrees of freedom it
would require to have $D$ independent constants of motion to keep a
regular motion (such systems are called integrable because, unlike 
what occurs in chaotic systems, in
principle the equations of motion can be solved at least 
locally in an appropriate
coordinate system).  This can be understood because the constants of
motion (energy, linear momentum, angular momentum) can be seen as
constraints imposing some restrictions on the dynamics: during its evolution, say, an isolated
system does not explore all the possible configurations, just the ones
that correspond to the same energy as the one it had initially. As
soon as a constant of motion is lost, for instance, by virtue of
Noether theorem, when a \textit{continuous} symmetry is broken (see
for instance \cite[\S~4]{Mouchet15a}), the system becomes chaotic.
The pendulum has $D=2$ degrees of freedom (say the coordinates~$(x,y)$
of the projection of its position on a horizontal plane).  The
regularity of the frictionless simple pendulum comes from the two
constants of motion: its energy (due to time-translation invariance)
and its angular momentum (due to rotational invariance with respect to
the vertical axis). On the contrary, even if friction were still
negligeable, the magnetic pendulum would still be chaotic precisely
because the external magnets break the continuous rotational
invariance\footnote{For the equilateral configuration considered in
  the previous section, there is still an invariance under a
  $120^\circ$ rotation, which is reflected in the patterns of
  figure~\ref{fig:pendmag_attracteur}e), but the invariance
  by a rotation of an
  \emph{arbitrary} angle is broken.}. Figure~\ref{fig:billards} provides another
illustration of the importance of symmetries for preserving a regular
motion.

\begin{wrapfigure}[13]{o}[0cm]{6cm}
\vspace{-.7cm}
\begin{center}
\includegraphics[width=5cm]{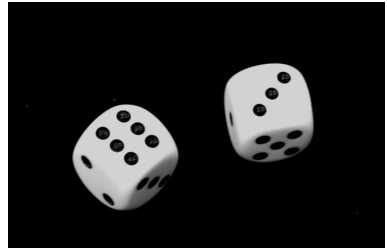}
\caption{\label{fig:des} It is the chaotic nature of the dynamics of rolling dices
that make them an appropriate random generator machine.   
}
\end{center}
\end{wrapfigure}
The lost of predictatibility is therefore ubiquitous and this leads to
the concepts of chance and contingency. 
Even if the dynamical equations are perfectly
deterministic, after a characteristic time~$\tau$, the system behaves
as if it were ``at random'', or more precisely what we call ``random''
or ``stochastic'' reveals a lost of the information that would 
be needed to predict the evolution of the system with no surprise.  From
what we have just discussed, it is therefore quite easy to get a
physical random generator, even more simpler and reliable than the
magnetic pendulum : a dice does the job or even the toss of a coin or a
lottery ball (figure~\ref{fig:des}).

\section{The stability of probabilities: the secret of the success of statistical physics}\label{sec:statistics}

The ubiquity of chaotic systems leads to two severe issues: (i) the
first concerns predictability which is one of the cornerstone of
science, (ii) the second concerns objectivity since the lack of
information that randomness represents depends a priori on the
observer: wouldn't a cleverer observer with more skills in modeling, more
ability of computing or more memory space find less randomness in
Nature? In 1814 Pierre-Simon de Laplace proposed to pursue this logic up
to an idealised intelligence for whom no chance would
exist\footnote{At that time, of course, the quantum indeterminism was
  not discovered yet.} \cite[p.~2]{Laplace1814}.

However, in practice, the exponential amplification of perturbations
during the evolution of a chaotic system simplifies the
situation. Dividing the initial uncertainty~$\delta\vec{r}(0)$ by~$10^n$
increases \textit{linearly} with~$n$, more precisely by~$n\tau\ln10$,
the time~$t$ when~$\delta\vec{r}$ reaches
a given value. So even if we take the ratio of the
size of the observable universe to the radius of the proton for which
$n$ is about 42, we extend the duration of reliability by less
than~$100\tau$ which is less than one minute for the magnetic
pendulum. Therefore from a physicist point of view, the time at which
a random behaviour appears is very robust and does not depend much on
the experimental conditions of the observation and can be safely
considered as objective. On the other hand, for times shorter
than~$\tau$ predictions with simple models remain reliable. The most
striking example, which is of first importance not only for having
contributed to the development of science but, above all, for having
provided a stable enough environment for intelligent life to evolve
on Earth, is provided by celestial mechanics. The two-body
gravitational problem leads to the regular Keplerian ellipses which
allow to reproduce the motion of the celestial bodies with a great
accuracy for centuries. However, as soon as a third body is involved
(not to speak of the eight main planets of the Solar System), there
cannot be found enough constants of motion to get a regular motion:
the discovery, by Henri Poincar\'e at the end of the
\textsc{xix}$^{\text{th}}$ Century, of what would be called chaos
actually comes from his mathematical studies on the motion of three
bodies interacting by gravity. As far as the Solar System is concern,
$\tau$ is of order of hundreds of millions years \cite{Laskar13a} and
predictability is safe at mankind scale (if we take into account
the small bodies like asteroids, that is another story).

These matters of time-scales are not the end of the argument of
course.  One way to answer to both issues (i) and (ii) is also to
consider collective effects and establish statistical laws. If the
predictability of one specific event or of one trajectory may truly be
challenged by a chaotic behaviour, the probability laws and some
collective properties obtained by averaging on many degrees of freedom
or on many observations provide reliable predictions and objective
facts. The origin of this stability can be traced back to the law of
large numbers: under general assumptions,
 the relative fluctuations within a sequence of~$N$
independent events are of order~$1/\sqrt{N}$.  For instance, whereas
we cannot predict the final position of the magnetic pendulum, we can
safely say that after~$N$ launching of the pendulum from various
equidistributed initial positions one will obtain about~$N/3$
trajectories ending on each~$M$ in the equilateral configuration and
this number is expected to fluctuate within a margin of
order~$\sqrt{N}$ as shown on the following table where the number of
coloured dots is extracted from the simulations in
figure~\ref{fig:pendmag_attracteur}.
\begin{center}
\begin{tabular}{|c|r||r|r|r|}
  \hline
   Grid of initial positions & $N/3$  &Red & Green & Blue \\
  \hline
   $10\times10$ & 33  & 36 & 36 & 28\\
   $25\times25$ & 208 & 220 & 218& 187 \\
   $50\times50$ & 833 & 815 & 827 & 858 \\
   $100\times100$ & 3\,333 & 3\,335 & 3\,321 & 3344 \\
   $500\times500$ & 83\,333 &83\,125 & 82\,965 &  83\,910 \\
  \hline
\end{tabular}
  \end{center}
A too large deviation would indicate that the configuration is biased and would indeed provide a quantitative measure of this bias. 

In statistical physics~$N$ is of the order of the Avogadro
number~$N_A\simeq6\ 10^{23}$ and, then, the relative fluctuations are of
order~$10^{-12}$ which is usually much below what is experimentally
accessible. Some quantities based on averages, like the temperature or
the pressure (figure~\ref{fig:emergencepression1}), are therefore very
relevant for describing physical properties at macroscopic scales.
 \begin{figure}[!ht]
\begin{center}
\includegraphics[width=15cm]{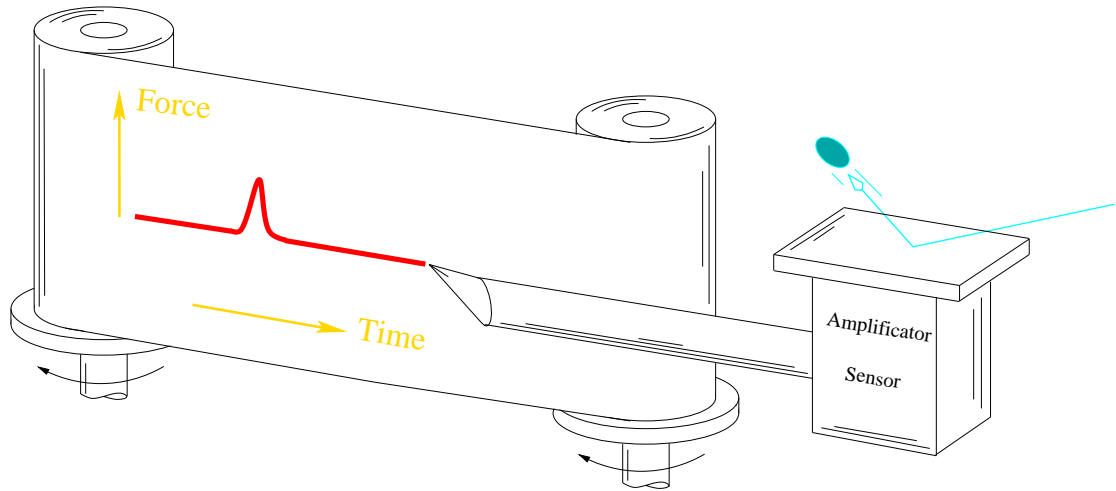} 
\caption{\label{fig:emergencepression0} A schematic device that is able to measure the recoil force
 due to one molecule hitting a plate.  
}
\end{center}
\end{figure}
\begin{figure}[!ht]
\begin{center}
\includegraphics[width=10cm]{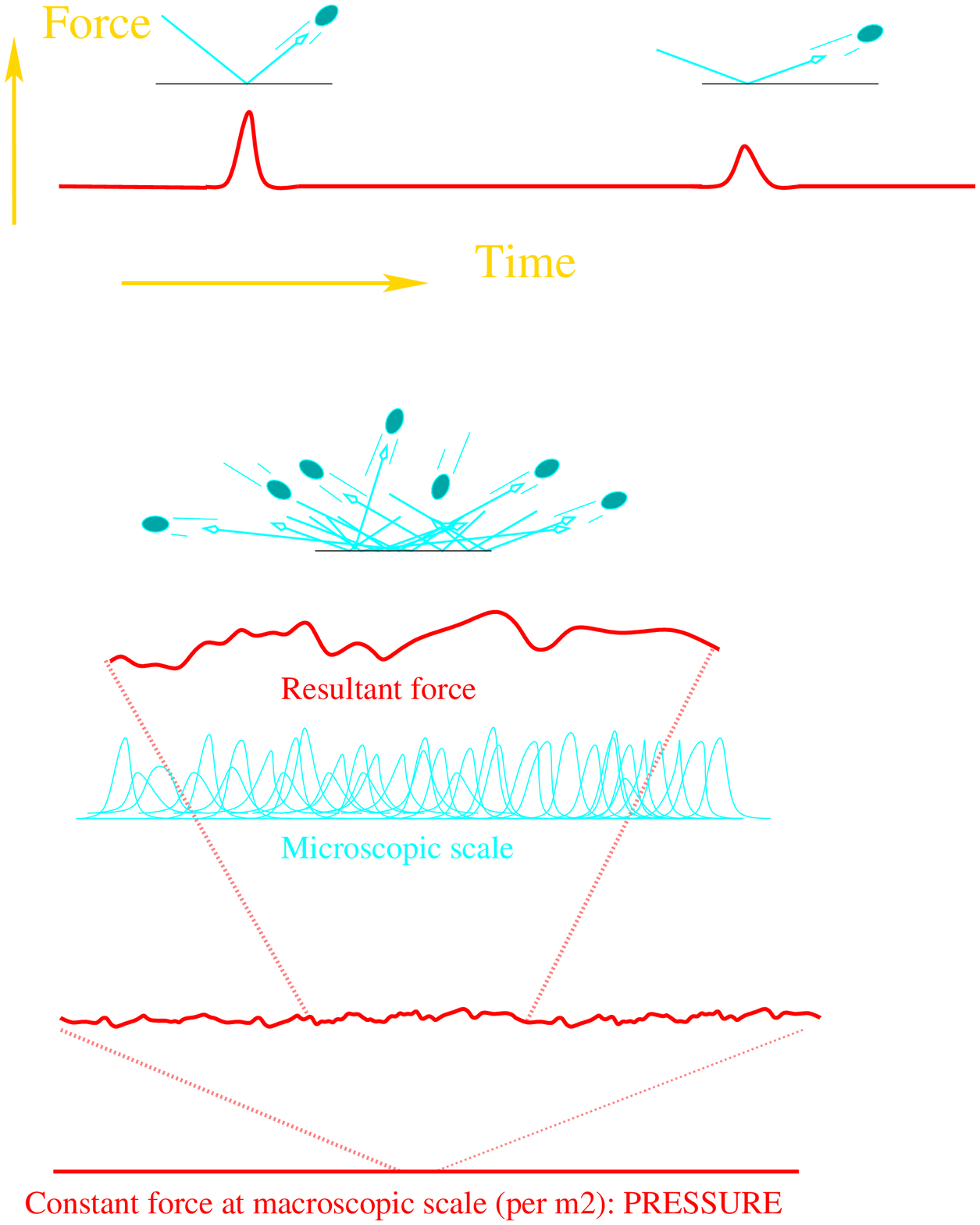}
\caption{\label{fig:emergencepression1} The device in
  figure~\ref{fig:emergencepression0} is used to measure the force on
  the plate when the number of molecules is increased. The more
  numerous the molecules, the more hits per unit of time and the less
  fluctuating the force is. At macroscopic scales, no fluctuations of
  the force can be detected and this constant average force per unit of
  surface is a stable quantity independent on the microscopic events:
  thus emerges the pressure.  }
\end{center}
\end{figure}
Although they are not well-defined at microscopic scales (talking of
the temperature or the pressure for one molecule is far-fetched),
 at larger scales such physical properties \emph{emerge} from a
collective effect. Although the statistical system is fully chaotic at
microscopic scales, the emergent properties at macroscopic scales are
insensitive to initial conditions. To comprehend a system one has to
give up almost all the information concerning its huge number of
parts: not only keeping this information
 is not feasible in practice but, above all, it
would not be relevant for a global study. For instance, all the
existing memory space (books, hard drive, etc.) could not register the
positions and the velocities of all the molecules in one litre of air;
had all these raw data be measurable and stored, they would be
completely over-abundant and would require to be synthesized to know,
say, if the container would resist when doubling the quantity of gas
inside it.  Abandoning almost all the information is not an
acknowledgement of weakness in front of complexity but rather a
necessity for bringing out the physical quantities that are pertinent
at large scales.  These emergent stable properties define most, if not
all, the macroscopic objects and are involved in the physical laws
governing them. Pruning the information from one level of a hierarchy
of physical structures (from quarks to cosmic filaments of
superclusters of galaxies, say) to the upper one helps to gain in
universality. The macroscopic quantities and the laws linking them become
independent of much more microscopic details than the position and the
velocities of its constituants.  The law of perfect gas~$PV=nRT$ is
independent of the nature of the constituents as long as their interactions are negligible: it remains valid for a
gas made of elements as different as one simple Helium atom and a much
more complex Carbon Dioxide molecule.  Another exemple of emergent property is
the transparency: it can be quantitatively defined for materials that,
at the microscopical level, are as different as water, diamond or
glass, although it is a non-sense of talking about the transparency of
an individual molecule.

This is all the art of statistical physics to identify the properties
that shoud be extracted from  a wide
collection of microscopic variables\footnote{In cite[\S\S~4.5.2, 4.6]{Mouchet13c}, I explain
how symmetries can be of great help in finding them.}. 
Climbing up the probably endless hierachy of physical systems is not less difficult or fundamental
than following the opposite way followed by the reductionnists.

\newpage
\section{A concluding historical comment}\label{sec:conclusion}

More than one hundred years ago Poincar\'e was aware that the extreme sensitivity  to initial conditions
would challenge meteorologists:

\begin{quotation}It may happen that 
small differences in the initial conditions produce very 
great ones in the final phenomena. A small error in 
the former will produce an enormous error in the 
latter. Prediction becomes impossible, and we have 
the fortuitous phenomenon. 
[\dots]\\
We will borrow [an example] from meteorology. Why 
have meteorologists such difficulty in predicting the 
weather with any certainty? Why is it that showers 
and even storms seem to come by chance, so that 
many people think it quite natural to pray for rain 
or fine weather, though they would consider it 
ridiculous to ask for an eclipse by prayer? We see 
that great disturbances are generally produced in 
regions where the atmosphere is in unstable equilibrium. 
The meteorologists see very well that the 
equilibrium is unstable, that a cyclone will be formed 
somewhere, but exactly where they are not in a 
position to say ; a tenth of a degree more or less at 
any given point, and the cyclone will burst here and 
not there. 
\cite[\S~II, p.~259-260]{Poincare07a}
\end{quotation}
\nocite{Poincare59a}

In the communication whose title was to give birth to the notion of
``butterfly effect'', it is often forgotten that, beyond the chaotic
behaviour of the Earth atmophere, Lorenz was putting forward the
stability of the statistics one may establish:

\begin{quotation}
More generally, I am proposing that over the years minuscule
disturbances neither increase nor decrease the frequency of
occurrences of various weather events such as tornados; the most they
may do is to modify the sequences in which they occur.  \cite[1st
  page]{Lorenz72a}
\end{quotation}
I hope the present contribution have respected Lorenz original message.

%\nocite{Lorenz93a}

%\bibliography{/home/mouchet/tex/localtexmf/bibtex/bst/mrabbrev,/home/mouchet/tex/localtexmf/bibtex/bst/bibliographie}

\begin{thebibliography}{10}

\bibitem{KosmannSchwarzbach10a}
Y.~{K}osmann{-S}chwarzbach.
\newblock \emph{The {N}oether theorems. Invariance and conservation laws in the
  twentieth century (sources and studies in the history of mathematics and
  physical sciences)}.
\newblock Springer, 2010.
\newblock Translated by Bertram E. Schwarzbach from the \ French original
  edition (Les {\'E}ditions de {l'\'E}cole Polytechnique, 2004).

\bibitem{Planck32a}
M.~Planck.
\newblock \emph{Where is science going?}
\newblock Norton \& Company, New York, 1932.
\newblock Translation by J. Murphy from the German \textit{Vom Relativen zum
  Absoluten} \cite{Planck25a}.

\bibitem{Einstein31a}
A.~Einstein.
\newblock \emph{Maxwell's influence on the development of the conception of
  physical reality}.
\newblock In \emph{James Clerk Maxwell, A Commemoration Volume 1831--1931}.
  Cambridge University Press, Cambridge, 1931, pages 66--73.

\bibitem{Mouchet15a}
A.~Mouchet.
\newblock \emph{Symmetry : a bridge between nature and culture}.
\newblock In Emmer et~al.  \cite{Emmer+15a}, pages 235--244.

\bibitem{Emmer+15a}
M.~Emmer, M.~Abate, and M.~Villarreal, editors.
\newblock \emph{Imagine Maths 4. Between culture and mathematics}. Unione
  Matematica Italiana. Instituto Veneto di Scienze, Lettere ed arti, Venezia e
  Bologna, 2015.

\bibitem{Pagel82a}
W.~Pagel.
\newblock \emph{Joan Baptista Van Helmont: Reformer of Science and Medicine}.
\newblock Cambridge Monographs in the History of Medicine. Cambridge University
  Press, Cambridge, 1982.

\bibitem{Lorenz72a}
E.~N. Lorenz.
\newblock \emph{Predictability: Does the flap of a butterfly's wings in
  {B}razil set off a tornado in {T}exas?} (1972).
\newblock Reproduced in \cite[Appendix~1]{Lorenz93a} and available at
  http://eaps4.mit.edu/research/Lorenz/Butterfly\_1972.pdf.

\bibitem{Mouchet13a}
A.~Mouchet.
\newblock \emph{L'{\'e}l{\'e}gante efficacit{\'e} des sym{\'e}tries}.
\newblock UniverSciences. Dunod, Paris, 2013.

\bibitem{Lorenz63a}
E.~N. Lorenz.
\newblock \emph{Deterministic nonperiodic flow}.
\newblock J. Atmospheric Sci., 20(2):pages 130--141 (1963).

\bibitem{Laplace1814}
P.-S. de~Laplace.
\newblock \emph{Essai philosophiques sur les probabilit\'es}.
\newblock Courcier, Paris, 1814.
\newblock English translation by F.~W. Truscott and F.~L. Emory \textit{A
  philosophical essay on probabilities}, Wiley, New York 1902. Laplace's
  complete works are available in French on http://gallica.bnf.fr/.

\bibitem{Laskar13a}
J.~Laskar.
\newblock \emph{Is the solar system stable?}
\newblock Progr. Math. Phys., 66:pages 239--270 (2013).
\newblock Available online arXiv:1209.5996.

\bibitem{Mouchet13c}
A.~Mouchet.
\newblock \emph{Reflections on the four facets of symmetry: how physics
  exemplifies rational thinking}.
\newblock Eur. Phys. J. H, 38:pages 661--702 (2013).

\bibitem{Poincare07a}
H.~Poincar{\'e}.
\newblock \emph{Le hasard}.
\newblock La Revue du Mois, 3:pages 257--276 (1907).
\newblock Available in French on
  http://henripoincarepapers.univ-lorraine.fr/bibliohp. Reproduced as chap.~IV
  of \cite{Poincare59a}.

\bibitem{Poincare59a}
H.~Poincar{\'e}.
\newblock \emph{Science and method}.
\newblock {D}over {P}ublications, {I}nc., {N}ew {Y}ork, 1959.
\newblock Translated by Francis Maitland from the French \textit{Science et
  m\'ethode} (Flammarion, 1908).

\bibitem{Lorenz93a}
E.~N. Lorenz.
\newblock \emph{The essence of chaos (Jessie and John Danz Lectures)}.
\newblock University of Washington Press, Washington, 1993.

\bibitem{Planck25a}
M.~Planck.
\newblock \emph{Vom relativen zum absoluten}.
\newblock Die Naturwissenschaften, 13(3):pages 53--59 (1925).
\newblock Translated in English in \cite{Planck32a}.

\end{thebibliography}

\end{document}